\documentclass[preprint,prb,
]{revtex4}

\usepackage{amsmath}

\newcommand{\bfc}{\boldsymbol{c}}

\newcommand{\gradv}{\boldsymbol{\nabla}}
\def\v#1{{\bf#1}}

\begin{document}

\title{The ${\bfc}$ equivalence principle and the correct \\form of writing Maxwell's equations}
\author{Jos\'e A. Heras}
\email{herasgomez@gmail.com}
\affiliation{Universidad Aut\'onoma Metropolitana Unidad Azcapotzalco, Av. San Pablo No. 180, Col. Reynosa, 02200, M\'exico D. F. M\'exico}

\begin{abstract}
\noindent 
It is well-known that the speed $c_u=1/\sqrt{\epsilon_0\mu_0}$ is obtained in the process of defining SI units via action-at-a-distance forces, like the force between two static charges and the force between two long and parallel currents. The speed $c_u$ is then physically different from the observed speed of propagation $c$ associated with electromagnetic waves in vacuum. However, repeated experiments have led to the numerical equality $c_u=c,$ which we have called the $c$ equivalence principle. In this paper we point out that $\gradv\times\v E=-[1/(\epsilon_0\mu_0 c^2)]\partial\v B/\partial t$ is the correct form of writing Faraday's law when the $c$ equivalence principle is not assumed. We also discuss the covariant form of Maxwell's equations without assuming the $c$ equivalence principle.
\end{abstract}

\maketitle
\noindent{\bf 1. Introduction}
\vskip 8pt
Textbooks on electromagnetism introduce the constant $\epsilon_0$ in electrostatics and the constant $\mu_0$ in magnetostatics. We would expect that the constant $1/\sqrt{\epsilon_0\mu_0}$ should also belong to the static regime of electromagnetic theory. However, when introducing the wave equation textbooks identify the quantity $1/\sqrt{\epsilon_0\mu_0}$ with $c$, the speed of propagation of time-dependent electric and magnetic fields which belong to the dynamical regime of the theory. Therefore the relation $1/\sqrt{\epsilon_0\mu_0}=c$ expresses a subtle connection between the static and dynamic regimes of electromagnetic theory. 
This connection is neither discussed in undergraduate nor graduate textbooks. In this paper we interpret the relation $1/\sqrt{\epsilon_0\mu_0}=c$ as a manifestation of the $c$ equivalence principle [1] expressed in SI units, which says that the speed $1/\sqrt{\epsilon_0\mu_0}$ emerging  from action-at-a-distance forces is equivalent to the speed $c$ of electromagnetic wave equations. We present some historical remarks related with this principle and write both the vector form  and the covariant form of Maxwell's equations when the $c$ equivalence principle is not assumed.
The discussion on the $c$ equivalence principle presented here is expected to be useful to understand the extension of the static regime of the electromagnetic theory to its dynamical regime and is intended to undergraduate and graduate students of electromagnetic theory.

\vskip 10pt
\noindent{\bf 2. Electrodynamics before Maxwell and the $\bfc$ equivalence principle}
\vskip 8pt
\noindent
The history of physics teaches us that the state of electromagnetic theory ``before Maxwell" was dominated by an instantaneous action-at-a-distance electromagnetic theory, which was represented by the set of equations (expressed in the modern SI notation) [2,3]:
\begin{align}
\gradv\cdot\v E&=\rho/\epsilon_0,\\
\gradv\cdot\v B&= 0,\\
\qquad\gradv\times \v E&=-\frac{\partial \v B}{\partial t},\\
\gradv\times \v B&=\mu_0\v J.
\end{align}
The constants $\epsilon_0$ and $\mu_0$ are seen to satisfy the relation
\begin{align}
c_u=\frac{1}{\sqrt{\epsilon_0\mu_0}}=2.998\times 10^5\;{\rm km/s}.
\end{align}
Before Maxwell the SI units were of course unknown. But let us use this little historical digression to be more pedagogical and modern our discussion. 

We first emphasize the action-at-a-distance origin of equation (5). By comparing the magnitude of the force between two static charges with the force between two long and parallel currents, we have recently derived the relation [1]:
\begin{align}
\frac{\alpha}{\beta\chi}=c_u^2,
\end{align}
where the constants $\alpha,\beta$ and $\chi$ are determined by the chosen units. Once these constants are defined, the value of the speed $c_u$ can be calculated. In SI units we chose $\beta=\mu_0=4\pi\times 10^{ -7}$ N/A$^2$ and $\chi=1$ and experimentally obtain: $\alpha=1/\epsilon_0$ with $\epsilon_0= 8.85\times 10^{-12}$ F/m [4]. Using these specific values for $\epsilon_0$ and $\mu_0$ we 
can directly verify the value of $c_u$ in equation~(5). Remarkably, equation (6) yields the same value of $c_u$ for other choice of units like Gaussian or Heaviside-Lorentz units. This means that the speed $c_u$ in equation (6) is independent of specific units and therefore it can be considered as a fundamental constant of nature [1]. 

The speed $c_u$ arises from using only action-at-a-distance forces in which an instantaneous propagation is assumed, or equivalently, where the speed of propagation is taken to be infinity. Therefore the finite speed $c_u$ cannot be associated with the instantaneous propagation of the fields $\v E$ and $\v B$ in equations~(1)-(4). However, in textbooks we find the equation  [5]: 
\begin{align}
c=\frac{1}{\sqrt{\epsilon_0\mu_0}}=2.998\times 10^5\;{\rm km/s},
\end{align}
where the speed $c$ is identified with the speed of light in vacuum. Equation (7) is usually introduced after deriving the wave equations for the electric and magnetic fields. It is pointed out that these fields propagate at the speed of light $c$. 

The identification of the speed $c_u$ with the speed $c$ should not be considered as an obvious result merely because these velocities
 have the same numerical value. The well-known example of the observed equality $m_g=m_i$ between the gravitational mass $m_g$ and the inertial mass $m_i$ of a body has taught us that numerical equivalence does not necessarily mean physical equivalence. The speeds $c_u$ and $c$ emerge from different physical considerations: $c_u$ is typical of action-at-a-distance laws which do not involve radiation and $c$ is typical of field theories which involve radiation. Since the speeds $c_u$ and $c$ are physically different but numerically equivalent, we have recently proposed that the equality [1]:
\begin{align}
c_u=c,
\end{align}
should be interpreted as the mathematical representation of the $c$ equivalence principle, just as $m_g\!=\!m_i$ formally represents the usual equivalence principle. The $c$ equivalence principle states that the speed $c_u$ emerging from action-at-a-distance electric and magnetic laws is equivalent to the propagation speed $c$ of electromagnetic waves in vacuum. Thus equation~(7) may be seen as a manifestation of the $c$ equivalence principle expressed in SI units:
\vskip 5pt
\centerline{\qquad{\it Action-at-a-distance\qquad\qquad Field action}\qquad\quad}
\vskip 10pt
\centerline{\boxed{\frac{1}{\sqrt{\epsilon_0\mu_0}}\;}\quad\qquad\;\;=\;\;\qquad\quad\boxed{\;c\;}}
\vskip 12pt
\noindent More in general, equation (8) may be interpreted as a manifestation of the $c$ equivalence principle expressed in a form independent of specific units: 
\vskip 12pt
\centerline{\boxed{\sqrt{\frac{\alpha}{\beta\chi}}\;}\quad\qquad\;\;=\;\;\qquad\quad\boxed{\;c\;}}
\vskip 10pt
\centerline{\qquad{\it Action-at-a-distance\qquad\qquad Field action}\qquad\quad}
\vskip 5pt
\noindent From a conceptual point of view the equality $c_u=c$ should be considered as an additional principle of the theory which unexpectedly 
links two completely different physical processes. It would not be an exaggeration to say that comparing action-at-a-distance and field action is like comparing apples and oranges!
Therefore, testing the $c$ equivalence principle to confirm its validity with very high precision would be an interesting and important task, just as the equivalence principle of gravitational and inertial masses is recurrently tested. 

The identification of the speed $c_u$ with the speed $c$  has sometimes caused astonishment and surprise. For example, Griffiths has pointed out [5]:  ``Remember how $\epsilon_0$ and $\mu_0$ came into the theory in the first place: there were constants in  Coulomb's law and Biot-Savart law, respectively. You measure them in experiments involving charged pith balls, batteries, and wires ---experiments having noting to do with light. And yet, according to Maxwell's theory you calculate $c$ from these two numbers." 

Most authors do not use two letters ($c_u$ and $c$) to identify the two different roles of the speed of light because they implicitly assume the $c$ equivalence principle. In a paper that discuss the different facets of $c$, Ellis and Uzan wrote [6]: ``Note that $c$ is not
only related to a velocity of propagation, because it can be measured by electrostatic and magnetostatic experiments." The Newtonian character of equation~(7) has been emphasized by  Preti {\it et al} [7] who wrote: ``the parameter $c$ in Maxwell's equations can actually be regarded as a property of Newtonian free space itself, due to its very definition [equation~(7)]
(SI units), in terms of two free space quantities, namely the vacuum permittivity $\epsilon_0$ and
the vacuum permeability $\mu_0$, which can be separately determined." These authors explain the basic motivation for the $c$ equivalence principle [7]:
``The fact that the numerical value of $c$ obtained
from equation [(7)] using these data [$\epsilon_0= 8.85\times 10^{-12}$ F/m and $\mu_0=4\pi\times 10^{ -7}$ N/A$^2$] actually coincides, within measurement precision, with the experimentally determined speed of propagation of light in vacuo, lets us usually speak of Maxwell's constant $c$ as `the speed of light
in vacuo' tout court." 

It follows that the standard form of writing Maxwell's equations in SI units is appropriate only if we adopt the $c$ equivalence principle in which case equation~(5) cannot be distinguished from equation~(7). The question naturally arises: How must Maxwell's equations be written when the $c$ equivalence principle is not assumed? This conceptually important question was briefly discussed in a recent paper [1]. It therefore seems appropriate to present a more detailed discussion of this topic
in the following sections.

\vskip 10pt
\noindent{\bf 3. Maxwell's equations}
\vskip 8pt
\noindent
The history of physics also teaches us that Maxwell realized that equation~(4) was only satisfactory for closed circuits.  He then generalized equation~(4) to an open circuit in a way consistent with the continuity equation by adding the term $\epsilon_0\mu_0 \partial \v E/\partial t$ to equation~(4). The resulting equations are now known as Maxwell's equations:
\begin{align}
\gradv\cdot\v E&=\rho/\epsilon_0,\\
\gradv\cdot\v B&= 0,\\
\qquad\gradv\times \v E&=-\frac{\partial \v B}{\partial t},\\
\gradv\times \v B&=\mu_0\v J+\epsilon_0\mu_0 \frac{\partial \v E}{\partial t}.
\end{align}
The most impressive prediction of equations~(9)-(12) are the wave equations:
\begin{align}
\nabla^2\v E-\frac{1}{c_u^2} \frac{\partial^2 \v E}{\partial t^2}&=\gradv\rho/\epsilon_0+\mu_0 \frac{\partial \v J}{\partial t},\\
\nabla^2\v B-\frac{1}{c_u^2} \frac{\partial^2 \v B}{\partial t^2}&=-\mu_0\gradv\times\v J,
\end{align}
where equation~(5) ---a legacy of equations (1)-(4)--- has been used. The direct interpretation of equations~(13) and (14) is universally accepted: these equations say that the fields $\v E$ and $\v B$ propagate (via an electromagnetic wave) from the sources $\rho$ and $\v J$ with the speed $c_u$, which is then ``naturally" identified with the speed of light $c$ in vacuum. What is wrong in this usual interpretation? Answer: The physical identification of the speed $c_u$ with the speed $c$. 
As above noted, the speeds $c_u$ and $c$ arise from different experimental considerations and have a distinct physical interpretation. 
Moreover, from historical point of view the symbol $c$ was originally introduced by Weber as a ratio of units of electric charge (this fact has  recently been pointed out by Mendelson in his review of the story of $c$ [9]). This ratio of units was measured later by Weber and Kohlrausch [10] who found the value $c_u= 3.1\times 10^5$ km/s. However, Weber did not see much of physical significance in the very approximate numerical coincidence between his value for $c_u$ and the value of the speed of light measured in that time. Kirchhoff [11] also noted the numerical coincidence between $c_u$ and the speed of light but, like Weber, he did not see such a coincidence as a possible hint about an electric theory of light [12]. 
Maxwell himself noted the physical differences of the velocities $c_u$ and $c$. In deriving one of the scalar components of the wave equation (in absence of sources), Maxwell [13] identified the velocity $c_u$ in that equation as determined by an experiment to measure the ratio of units$^{\ast}$ 
\footnotetext[1]{In esu units we write $\alpha=4\pi$ and experimentally obtain $\beta\chi=4\pi/c_u^2$ and chose $\beta=4\pi/c_u^2$ and $\chi=1$. In emu units we chose $\beta\chi=4\pi$ with $\beta=4\pi$ and $\chi=1$ and experimentally obtain $\alpha=4\pi c_u^2$. If $q_{esu}$ and  $q_{emu}$ denote a charge in emu and esu units then  $F\!=\!q_{esu}^2/R^2$ and $F\!=\!c_u^2q_{emu}^2/R^2$ from which we obtain
\begin{align}
c_u=\frac{q_{esu}}{q_{emu}}.\nonumber
\end{align}
This relation means that the speed $c_u$ can be defined as a ratio of units of charge. Maxwell wrote: ``$[c_u]$... is the number by which the electrodynamic measure of any quantity of electricity must be multiplied to obtain its electrostatic measure." See Siegel D M 1991 {\it Innovations in Maxwell's electromagnetic theory: Molecular Vortices, Displacement Current,
and Light} (Cambridge: Cambridge University Press) p 82}without any intervention of the speed of light $c$. He wrote a nice phrase [13]: ``The only use made of light in this experiment was to see the instruments[!]" In Maxwell's times the speed of light was considered to be a result of purely optical origin. Furthermore, Maxwell examined the experiment of Foucault to measure the speed of light $c$ by optical means and concluded that in this experiment [14]: ``No use whatever was made of electricity or magnetism." 

In performing the generalization of equations~(1)-(4) to equations~(9)-(12) the speed $c_u$ has been implicitly inherited (through the quantities $\epsilon_0$ and $\mu_0$) to the latter set of equations. The traditional interpretation of equations~(13) and (14) has then changed the meaning of the speed $c_u$ with no physical justification, by simply stating that the speed $c_u$ is now the speed of propagation $c$ of the fields $\v E$ and $\v B$. 
\vskip 10pt
\noindent{\bf 4. Extending Jammer and Stachel's fable}
\vskip 8pt
\noindent
We can answer the question of how Maxwell's equations must be written when $c_u=c$ is not assumed, by extending Jammer and Stachel's historical fable [8] in which the course of history has been reconstructed by supposing that Maxwell had been working before Faraday had discovered his law. Let us briefly recreate this fable. It seems to be natural that Maxwell would have generalized the Coulomb and Ampere static equations to directly include time dependence: $\gradv\cdot\v E(\v x,t)=\rho(\v x,t)/\epsilon_0,\gradv\times \v E(\v x,t)= 0$, $\gradv\cdot\v B(\v x,t)= 0 $, and $\gradv\times \v B(\v x,t)=\mu_0\v J(\v x,t)$. He would have surely noticed that these equations were inconsistent with the continuity equation. To remedy this defect he could have introduced the term $\epsilon_0\mu_0\partial \v E/\partial t$
into the curl of $\v B$ obtaining the field equations of an instantaneous action-at-a-distance theory, which are shown to be Galilei-invariant [8]:
\begin{align}
\gradv\cdot\v E&=\rho/\epsilon_0,\\
\gradv\cdot\v B&= 0,\\
\gradv\times \v E&=0,\\
\gradv\times \v B&=\mu_0\v J+\epsilon_0\mu_0 \frac{\partial \v E}{\partial t}.
\end{align}
These equations, imaginarily discovered by Maxwell, would describe the state of electromagnetic theory ``before Faraday." Faraday then could have appeared on the scene to finish the work by introducing the term $-\partial\v B/\partial t$ into the right-hand side of equation~(17) and in this way he would have arrived at equations~(9)-(12) which break Galilean invariance but acquire Lorentz invariance. End of Jammer and Stachel's fable [8].

Let us extend the final part of the fable when Faraday would have begun his work having equations~(15)-(18) on the table. Faraday
was a good experimentalist. Let us suppose that he was a good mathematician as well.  Then he might have discovered that time variations of the field $\v B$ are connected with the field $\v E$ and therefore he would have surely concluded that equation~(17) had necessarily to be modified. Such a modification would be of the form 
\begin{align}
\gradv\times \v E=k\frac{\partial \v B}{\partial t},
\end{align}
where $k$ is a constant to be determined. To decouple the field $\v B$ in equations~(16), (18) and (19), Faraday might have combined these equations obtaining the differential equation
\begin{align}
\nabla^2\v B +\epsilon_0\mu_0 k\frac{\partial^2\v B}{\partial t^2}=-\mu_0\gradv\times\v J.
\end{align}
The next step would have been crucial to Faraday. The mathematical character of equation~(20) was even undefined. Following his initial idea that the physical reality was in the fields (according to him action-at-a-distance was not the best picture for electromagnetic phenomena) and, in addition, assuming that they propagate at the finite speed $c$, Faraday might have then concluded that equation~(20) should be a wave equation   
for the field $\v B$ with the propagation speed $c$.  This conclusion would demand the validity of the relation $\epsilon_0\mu_0 k=-1/c^2$, from which $k$ is determined: $k=-1/(\epsilon_0\mu_0 c^2)$. With this value of $k$, Faraday would have obtained the required modification of equation~(17):
$\gradv\times\v E=-[1/(\epsilon_0\mu_0 c^2)]\partial\v B/\partial t$ as well as the wave equation:
\begin{align}
\nabla^2\v B -\frac{1}{c^2}\frac{\partial^2\v B}{\partial t^2}=-\mu_0\gradv\times\v J.
\end{align}
At the end of his work, Faraday would have generalized equation~(15)-(18) obtaining a set of equations that might have been called  ``Faraday's equations":
\begin{align}
\gradv\cdot\v E&=\rho/\epsilon_0,\\
\gradv\times \v E&=-\frac{1}{\epsilon_0\mu_0 c^2}\frac{\partial \v B}{\partial t},\\
\gradv\cdot\v B&= 0,\\
\gradv\times \v B&=\mu_0\v J+\epsilon_0\mu_0 \frac{\partial \v E}{\partial t}.
\end{align}
By combining equations~(22)-(25) Faraday might have verified that the field $\v E$ also satisfies a wave equation with the speed of propagation $c$: 
\begin{align}
\nabla^2\v E -\frac{1}{c^2}\frac{\partial^2\v E}{\partial t^2}=\gradv\rho/\epsilon_0+\mu_0\frac{\partial\v J}{\partial t}.
\end{align}
Let us go even further in the fable by imagining that Maxwell re-appeared on the scene when Faraday's equations~(22)-(25) were already on the table. Maxwell then might have compared the value of $c_u=1/\sqrt{\epsilon_0\mu_0}$ calculated on purely electromagnetic considerations with the value of the speed of light $c$ as measured by purely optical means. The agreement of these two values would have lead him to claim that $c_u=c$.$^{\dagger}$ \footnotetext[2]{Maxwell wrote: ``The agreement of the results $[c_u$ and $c$ ] seems to shew that light and magnetism are affections of the same substance, and that light is an electromagnetic disturbance propagated through the field according to electromagnetic laws." 
See Siegel D M 1991 {\it Innovations in Maxwell's electromagnetic theory: Molecular Vortices, Displacement Current,
and Light} (Cambridge: Cambridge University Press) p 155} He then immediately would have noted the result  $1/(\epsilon_0\mu_0 c^2)=1$ which implies that Faraday's equations (22)-(25) become Maxwell's equations (9)-(12). He then might have concluded that [15]: ``...our theory, which asserts that these two quantities $[c_u$ and $c]$ are equal, and assigns a physical reason for this equality [light is an electromagnetic wave] is certainly not contradicted by comparison of [experimental] results." Perhaps this was the first evocation of the $c$ equivalence principle.  

We end the extended fable and answer the question posed at the end of the second section, by claiming that equations~(22)-(25) are the correct form of writing Maxwell's equations without assuming the validity of the $c$ equivalence principle. If one explicitly assumes this principle then equations~(9)-(12) are correctly expressed as they stand. Incidentally, it has been shown [1] that equations~(22)-(25) [and not equations~(9)-(12)]  are the appropriate form to obtain the instantaneous limit of Maxwell's equations given by equations~(15)-(18).  In fact, this limit is obtained by letting $c\to\infty$ into equations~(22)-(25) and keeping $c_u$ intact (this limit actually implies a violation of the $c$ equivalence principle). For completeness we write the form of equations~(22)-(25) in Gaussian units in Appendix A. 
\vskip 10pt
\noindent{\bf 5. Covariant form of Maxwell's equations without the $\bfc$ equivalence principle}
\vskip 10pt
\noindent
We proceed now to show how the covariant form of Maxwell's equations in SI units can be written without assuming the $c$ equivalence principle. We will formulate equations~(22)-(25) in Minkowski spacetime.
The expected tensor equations will exhibit the same form than the usual covariant form of Maxwell's equations but with a different definition of the electromagnetic field tensor $F^{\mu\nu}.$$^{\ddagger}$\footnotetext[3]{ 
Greek indices $\mu, \nu, \kappa ...$ run from 0 to 3; Latin indices $i,j,k,...$ run from 1 to 3;
$x=x^{\mu}=(x^0,x^i)=(ct,\v x)$ is the field point and $x'=x'^{\mu}=(x'^0,x'^i)=(ct',\v x')$ the source point; 
the signature of the metric $\eta^{\mu\nu}$ of the Minkowski spacetime  is $(+,-,-,-);$ $\varepsilon^{\mu\nu\alpha\beta}$ is the 
totally antisymmetric four-dimensional tensor with $\varepsilon^{0123}=1$ and $\varepsilon^{ijk}$ is the totally antisymmetric three-dimensional tensor with $\varepsilon^{123} = 1.$ Summation convention on repeated indices is adopted. A four-vector in spacetime can be represented in the so-called (1+3) notation as $F_\nu=(f_0,\v F),$ where $f_0$ is its time component and $\v F$ its space component.
Derivatives in spacetime are defined by $\partial_\mu=[(1/c)\partial/\partial t, \gradv]$ and  $\partial^\mu=[(1/c)\partial/\partial t,- \gradv].$}
In fact, the covariant form of equations~(22)-(25) reads
\begin{align}
\partial_\mu F^{\mu\nu} &= \mu_0J^\nu, \\
\partial_\mu\!{^*}\!F^{\mu\nu} &=  0,
\end{align}
where $J^\nu$ is the four-current given by
\begin{align}
J^\mu=(c\rho,\v J),
\end{align}
and $^*\!F^{\mu\nu}\!=\!(1/2)\varepsilon^{\mu\nu\kappa\sigma}\!F_{\kappa\sigma}$ is the dual of the tensor $F^{\mu\nu}$\!.
The components of $F^{\mu\nu}$ are given by 
\begin{align}
F^{i0}&=\mu_0\epsilon_0 c(\v E)^i, \\
F^{ij}&=-\varepsilon^{ijk}(\v B)_k.
\end{align}
where $(\v E)^i$ and $(\v B)_k$ represent the scalar components of the electric and magnetic fields. The components of $^*\!F^{\mu\nu}$ can be obtained from those of $F^{\mu\nu}$ by making the dual changes: 
\begin{align}
\mu_0\epsilon_0 c(\v E)^i&\rightarrow (\v B)^i,\\
(\v B)_k &\rightarrow -\mu_0\epsilon_0 c(\v E)_k. 
\end{align}
Therefore the components of  ${^*}\!F^{\mu\nu}$ are given by
\begin{align}
^*\!F^{i0}=&(\v B)^i,\\
^*\!F^{ij}=&\mu_0\epsilon_0 c\varepsilon^{ijk}(\v E)_k.
\end{align}
With the aid of the above definitions, we can write the following four-vectors:
\begin{align}
\partial_{\mu}F^{\mu\nu}=&\bigg (\mu_0\epsilon_0 c\nabla \cdot \v E,\;\nabla \times \v B-\mu_0\epsilon_0\frac{\partial \v E}{\partial t} \bigg ),\\
\partial_{\mu}\!{^*}\!F^{\mu\nu}=&\bigg (\nabla \cdot \v B, \; -\mu_0\epsilon_0c\nabla \times \v E-\frac 1c\frac{\partial \v B}{\partial t} \bigg) .
\end{align}
To obtain equations~(22) and (25) we make equal the time and space components in both sides of equations~(27) and (29) and use equation~(36). Next we make zero the time and space components of equations~(28) and use equation~(37) to obtain equations~(23) and (24). If we assume $c_u=c$ then $\epsilon_0\mu_0=1/c^2$ and therefore equations~(27) and (28) become the standard covariant form of Maxwell's equations in SI units [16].
In Appendix A we write equations~(27) and (28) in Gaussian units.
\vskip 10pt
\noindent{\bf 6. Summary}
\vskip 8pt
\noindent
Historical facts are not necessarily the best pedagogical tools to understand a theory. Jammer and Stachel's approach [8], which is drawn as a fable that reverses historical findings by introducing first the displacement current term in quasistatic forms of Maxwell's equations before introducing  Faraday's  induction term, is a useful and pedagogical alternative to introduce Maxwell's equations. However, Jammer and Stachel's fable does not discuss in detail how Faraday introduced his term to obtain the final equations.

We have extended here Jammer and Stachel's fable [8] by imagining how Faraday's term could have been introduced considering that 
the speed $c_u=1/\sqrt{\epsilon_0\mu_0}$ (a legacy of action-at-a-distance laws) is physically different from the speed of light $c$ associated with electromagnetic waves. We have called attention that the observed equality $c_u=c$, which we have called the $c$ equivalence principle [1], is a conceptually important relation which should be considered to be an additional axiom of the theory and noted that this equality was first emphasized by Maxwell [15]. Without assuming this equality, Faraday's law must be written as shown in equation~(23). The form of equations~(9)-(12) is correct if one assumes the $c$ equivalence principle. Otherwise, the correct form is given by equations~(22)-(25). We have also discussed the covariant form of Maxwell's equations without assuming the $c$ equivalence principle.

\appendix 
\section{}\noindent{\bf Gaussian units}
\vskip 6pt
\noindent Using equation~(6) we can define Gaussian units. We specify $\alpha=4\pi$ and experimentally obtain $\beta\chi=4\pi/c_u^2$. For these units we chose $\beta=4\pi/c_u$ and $\chi=1/c_u$. Maxwell's equations in Gaussian units without assuming the $c$ equivalence principle are given by [1]:
\begin{align}
\gradv\cdot\v E&=4\pi\rho,\\
\gradv\cdot\v B&= 0,\\
\qquad\gradv\times \v E&=-\frac{c_u}{c^2}\frac{\partial \v B}{\partial t},\\
\gradv\times \v B&=\frac{4\pi}{c_u}\v J+ \frac{1}{c_u}\frac{\partial \v E}{\partial t}
\end{align}
Evidently, if $c_u=c$ then we recover the usual form of these equations. 

The covariant form of equations~(A1)-(A4) is given by 
\begin{align}
\partial_\mu F^{\mu\nu} &= \frac{4\pi}{c_u}J^\nu, \\
\partial_\mu\!{^*}\!F^{\mu\nu} &=  0,
\end{align}
where the components of $F^{\mu\nu}$ are given by
\begin{align}
F^{i0}&=\frac{ c}{c_u}(\v E)^i, \\
F^{ij}&=-\varepsilon^{ijk}(\v B)_k.
\end{align}
and the components of its dual tensor $^*\!F^{\mu\nu}$ by
\begin{align}
^*\!F^{i0}=&(\v B)^i,\\
^*\!F^{ij}=&\frac{ c}{c_u}\varepsilon^{ijk}(\v E)_k.
\end{align}
The four vectors $\partial_{\mu}F^{\mu\nu}$ and $\partial_{\mu}\!{^*}\!F^{\mu\nu}$ are defined by 
\begin{align}
\partial_{\mu}F^{\mu\nu}=&\bigg( \frac{c}{c_u}\nabla \cdot \v E,\;\nabla \times \v B-\frac{1}{c_u}\frac{\partial \v E}{\partial t} \bigg ),\\
\partial_{\mu}\!{^*}\!F^{\mu\nu}=&\bigg (\nabla \cdot \v B, \; -\frac{c}{c_u}\nabla \times \v E-\frac 1c\frac{\partial \v B}{\partial t} \bigg) .
\end{align}
If $c_u=c$ then we recover the covariant form of Maxwell's equations in Gaussian units [16].

{}

\end{document}